\documentclass[aps,prl,twocolumn,superscriptaddress,amsfonts,amssymb]{revtex4}

\usepackage{graphicx}
\usepackage{dcolumn}

\begin{document}


\title{NMR and Electronic Structures in Type-I BaAlGe Clathrates}


\author{Weiping Gou}
\author{Sergio Y. Rodriguez}
\affiliation{Department of Physics, Texas A\&M University, College
Station, TX 77843-4242}
\author{Yang Li}
\affiliation{Dept. Engineering Science and Materials, Univ. Puerto Rico at Mayaguez,
Mayaguez, PR  00681-9044}
\author{Joseph H. Ross, Jr.}
\affiliation{Department of Physics, Texas A\&M University, College
Station, TX 77843-4242}


\date{\today}

\begin{abstract}
We describe $^{27}$Al NMR experiments on Ba$_{8}$Al$_{x}$Ge$_{46-x}$ type-I clathrates coupled with {\it ab initio} computational studies. For 
$x$ = 16, calculated spectra determined by the {\it ab initio} results gave good agreement with the measurements, with best-fitting configurations also corresponding to the computed lowest-energy atomic arrangements. Analysis of the NMR results showed that a distribution of Knight shifts dominates the central portion of the line. Computational results demonstrate that this stems from the large variation of carrier density on different sites. Al-deficient samples with $x$ = 12 and 13 exhibited a split central NMR peak, signaling two main local environments for Al ions, which we connected to the presence of vacancies. Modeling of the wide-line spectrum for $x$ = 12 indicates a configuration with more Al on the 24$k$ site than for $x$ = 16. The results indicate the importance of non-bonding hybrids adjacent to the vacancies in the electronic structure near $E_{F}$. We also address the static distortions from $Pm\overline{3}n$ symmetry in these structures.
\end{abstract}

\maketitle

\section {INTRODUCTION}

In recent years clathrates have gained considerable attention due to their thermal and thermoelectric properties, and potential for device application \cite{Nolas98,Cohn99,Kuznetsov00,Chakoumakos00,Ross09}. Clathrates consist of cages of silicon, germanium, or tin in a crystalline framework, with guest atoms located inside the cages. The type-I clathrates have the general formula $M_{8}X_{46}$, where $\textit{M}$ and $\textit{X}$ represent the guest atoms and framework atoms, respectively. Each unit cell contains two 20-atom cages which form pentagonal dodecahedra, and six 24-atom cages which form tetrakaidecahedra. Recently much work has been devoted to Ba$_{8}$Ga$_{16}$Ge$_{30}$, which exhibits a particularly large thermoelectric efficiency at high temperatures \cite{Saramat06}. An important issue has been to understand the displacement of Ba ions inside the larger cages, and the strong resulting effect on transport behavior.

It was recently shown that Al alloying in Ba$_{8}$Ga$_{16-x}$Al$_{x}$Ge$_{30}$ can act to tune the thermal conductivity \cite{deng08}, however it was also shown by high-resolution crystallography that in Ba$_{8}$Al$_{16}$Ge$_{30}$ the interactions between Ba and group-III Al ions strongly affect the vibrational behavior of cage-center ions \cite{Christensen07}. An additional feature which strongly affects the electronic behavior is the relative ease of breaking the Ge framework, with vacancies spontaneously formed to maintain electron balance in off-stoichiometry materials. Thus it is interesting to examine the electronic behavior of these materials using a local probe spectroscopy. In this paper, we report a study of Ba$_{8}$Al$_{x}$Ge$_{46-x}$ with $x$ = 12--16, examining local variations in electronic behavior as the composition is varied and vacancies are formed.  

\section{ SAMPLE PREPARATION AND CHARACTERIZATION}

Samples of nominal composition Ba$_{8}$Al$_{x}$Ge$_{46-x}$ were prepared from the elements, mixed according to the desired composition, by arc melting in argon following by annealing at 500 $^{\circ}$C for two days. Samples are denoted by Al$_{x}$, for example Al$_{16}$ representing 
Ba$_{8}$Al$_{16}$Ge$_{30}$. Powder x-ray diffraction measurements showed the Al$_{13}$ and Al$_{16}$ samples to be pure type-I clathrate 
with no additional phases within experimental resolution, while Al$_{12}$ and Al$_{10}$ also exhibited a small amount of diamond-phase Ge. Rietveld analysis yielded best fits of Ba$_{8.0}$Al$_{15.1}$Ge$_{27.8}$ for the nominal Al$_{16}$ sample, and 
Ba$_{7.0}$Al$_{12.6}$Ge$_{33.5}$ for Al$_{12}$. For Al$_{16}$ these yielded 2.0, 5.2, and 7.9 Al atoms on the 6$c$, 16$i$, and 24$k$ framework sites respectively in the type-I structure (space group $Pm\overline{3}n$, \#223), and for Al$_{12}$, 1.5, 2.0, and 9.1 Al atoms.
Lattice constants gradually decreased from 10.852 {\AA} for Al$_{16}$  to 10.838 {\AA} for Al$_{12}$ and 10.814 {\AA} for Al$_{10}$. Resistivities exhibited positive temperature coefficients, typical of metals or heavily-doped semiconductors, with, for example, Al$_{16}$ having a low-temperature value of 170 $\mu \Omega$-cm, within the range observed previously for Ba$_{8}$Al$_{16}$Ge$_{30}$ samples \cite{Christensen07}. The Al$_{12}$ susceptibility is diamagnetic, with a temperature-independent value of 
$- 1.6 \times 10^{-3}$ emu/mol; similar to other Ge clathrates \cite{Li03} this is larger than the summed core diamagnetic contributions ($- 0.5 \times 10^{-3}$ in this case).
 
A feature of Ba$_{8}$Al$_{x}$Ge$_{46-y}$ is the occurrence of spontaneous vacancies to balance the electron count in the framework, according to the Zintl concept \cite{Miller96}. For example, for Al$_{12}$ the nominal Zintl composition is 
Ba$_{8}$Al$_{12}$Ge$_{33}\Box_1$ ($\Box$ indicating a vacancy), since if each Ba donates two electrons and Al has valence 3, there are four excess framework electrons per cell after filling all bonds, enough to fill the nonbonding orbitals adjoining a single vacancy. The expected result is a semiconductor or low-carrier density material.
Ba$_{8}$Al$_{16}$Ge$_{30}$ has a balanced framework electron count and no spontaneous vacancies, while as the Al content decreases the vacancy count would increase, the limiting case being Ba$_{8}$Ge$_{43}\Box_3$, for which the vacancies form an ordered superstructure \cite{Fukuoka03,Budnyk04,Shimizua07}.
Such behavior is also seen in 
Ba$_{8}$Al$_{16-x}$Si$_{30+y}$ \cite{Mudryk02,Condron06}, however vacancies form less readily in the Si framework, for example Ba$_{8}$Si$_{46}$ 
remaining a strongly metallic superconductor \cite{Yamanaka00,Ross09}.
For Ba$_{8}$Ga$_{x}$Ge$_{46-y}$ the vacancy count was determined to be about half the expected Zintl value over a range of compositions \cite{Okamoto06}, while scans of our nominal Al$_{10}$ sample by wavelength dispersion spectroscopy (WDS, an electron microprobe method), yielded a uniform clathrate phase with 
composition Ba$_{8}$Al$_{9.2}$Ge$_{33.6}$, thus indicating three vacancies per formula unit, similar to the number present in 
Ba$_{8}$Ge$_{43}$, and somewhat larger than the expected number for a Zintl phase. We saw no superstructure peaks in powder XRD that might indicate long-range vacancy ordering as seen in Ba$_{8}$Ge$_{43}$. 

NMR experiments were performed at a field of 8.9 T using a pulse spectrometer described previously \cite{lue98}. $^{27}$Al NMR shifts were calibrated using dilute aqueous AlCl$_{3}$ as zero-shift reference. Samples were powdered and mixed with KBr as an insulating medium. Magic Angle Spinning (MAS) measurements were performed at room temperature using a Bruker Avance 400 MHz NMR instrument, with Al$_{2}$O$_{3}$ as reference. 

\section{NMR MEASUREMENTS}

\begin{figure}
\includegraphics[width=\columnwidth]{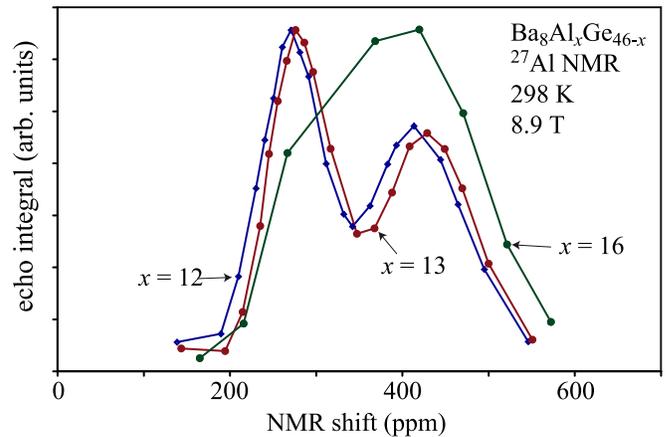}
\caption{\label{fig:fig1} (Color online) NMR central lines for Ba$_{8}$Al$_{x}$Ge$_{46-x}$, $x$ = 12, 13, and 16, at room temperature.}
\end{figure}

\begin{figure}
\includegraphics[width=\columnwidth ]{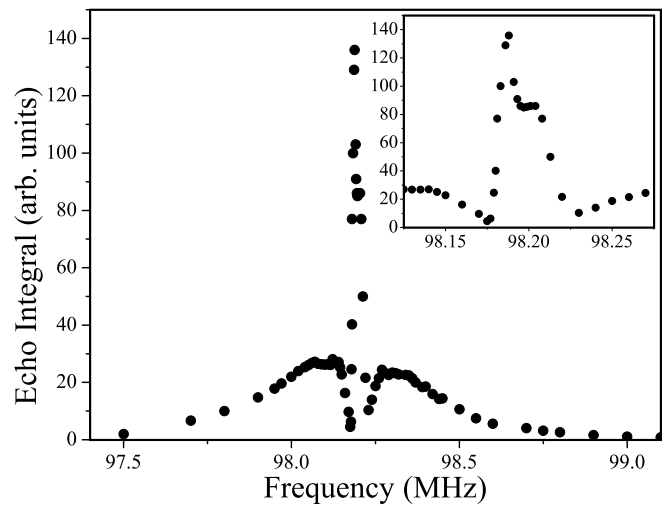}
\caption{\label{fig:fig5}Full NMR lineshape for Ba$_{8}$Al$_{12}$Ge$_{34}$ from a spin-echo measurement at 4.2 K. Inset shows central line portion of the spectrum.}
\end{figure}

NMR spectra of samples Al$_{12}$, Al$_{13}$, and Al$_{16}$ measured by spin echo
integration at room temperature are shown in Fig.~\ref{fig:fig1}. For Al$_{12}$ and Al$_{13}$ we observe two obvious peaks, 
while for Al$_{16}$ a single, broader line is seen. These central lines are surrounded by a more extended resonance signal, mapped out in detail at lower temperatures, as shown in
Fig.~\ref{fig:fig5} for Al$_{12}$. The central lines correspond to $\Delta m$ = 1/2 to --1/2 transitions for the $I$ = 5/2 $^{27}$Al nucleus, while the broad background is due to a superposition of the other four transitions. The two peaks in the Al$_{12}$ central line overlap more strongly in 
Fig.~\ref{fig:fig5} than in Fig.~\ref{fig:fig1} due to relative changes in NMR shift vs. temperature. The valleys covering about 50 kHz on either side of the central line are artifacts due to differences in pulse response for the five transitions. For the type-I structure, the three framework sites all have some Al occupation as indicated by the x-ray fits, and in addition there are distinct local neighbor configurations due to the varied occupation of these sites. Thus the NMR lines are superpositions of signals from multiple local environments, with their individual powder patterns. 

\begin{figure}
\includegraphics[width=2.8 in]{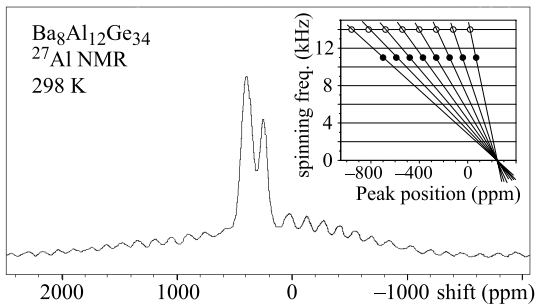}
\caption{\label{fig:fig7}Spinning NMR measurement of Ba$_{8}$Al$_{12}$Ge$_{34}$ at room temperature, with 14 kHz rotor speed, and higher frequencies to the left. Inset: rotor speed dependence shows that broad background corresponds to lower-shift peak. Note reversal of paramagnetic shift axis orientation relative to other figures.}
\end{figure}

A spinning spectrum for Al$_{12}$ is shown in Fig.~\ref{fig:fig7}. Magic Angle Spinning (MAS) can narrow first-order quadrupole powder patterns such as the broad satellite lines seen here \cite{Slichter}. The Al$_{12}$ MAS spectrum has a two-peaked central line like that of the static spectrum, flanked by spinning sidebands in place of the near portions of the non-central resonances. Note that the paramagnetic shifts increase to the left in this figure, the reverse of the orientation in Fig.~\ref{fig:fig1}. The further wings of the quadrupole powder pattern appear in the spin-echo measurements (Fig.~\ref{fig:fig5}) but fall out of range of the MAS measurement. Al$_{13}$ yielded a similar result, though with sidebands less distinct indicating increased inhomogeneous broadening. The spinning frequency dependence of the sidebands (inset, Fig.~\ref{fig:fig7}), showed the center to be 240 ppm, underneath the low frequency peak. Thus, the near portions of the non-central lines are associated with this part of the central line spectrum. For Al$_{16}$ MAS revealed a broadened central-line like that of the static spectrum, again flanked by spinning sidebands. A spinning-frequency analysis (not shown) showed the non-central background to be centered at 330 ppm, near the center of the featureless central line for this sample, and somewhat higher than for Al$_{12}$. 

\begin{figure}
\includegraphics[width=\columnwidth]{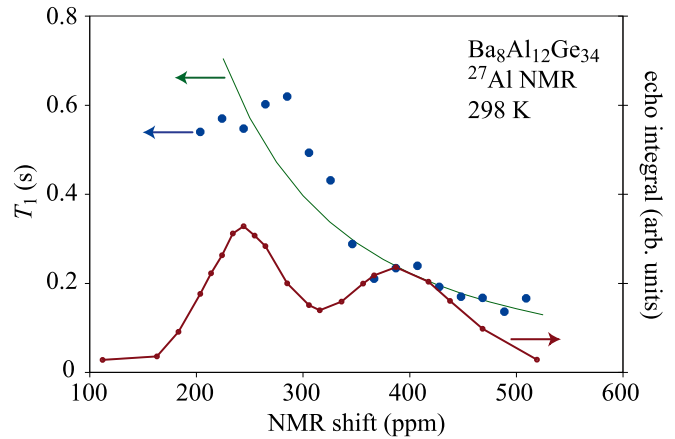}
\caption{\label{fig:fig2}(Color online) Frequency dependence of $T_{1}$ for Ba$_{8}$Al$_{12}$Ge$_{34}$ compared to central line shape (lower curve). Solid curve with monatonic decay is $T_{1}$ calculated according to Korringa law, described in text.}
\end{figure}

\begin{figure}
\includegraphics[width=\columnwidth]{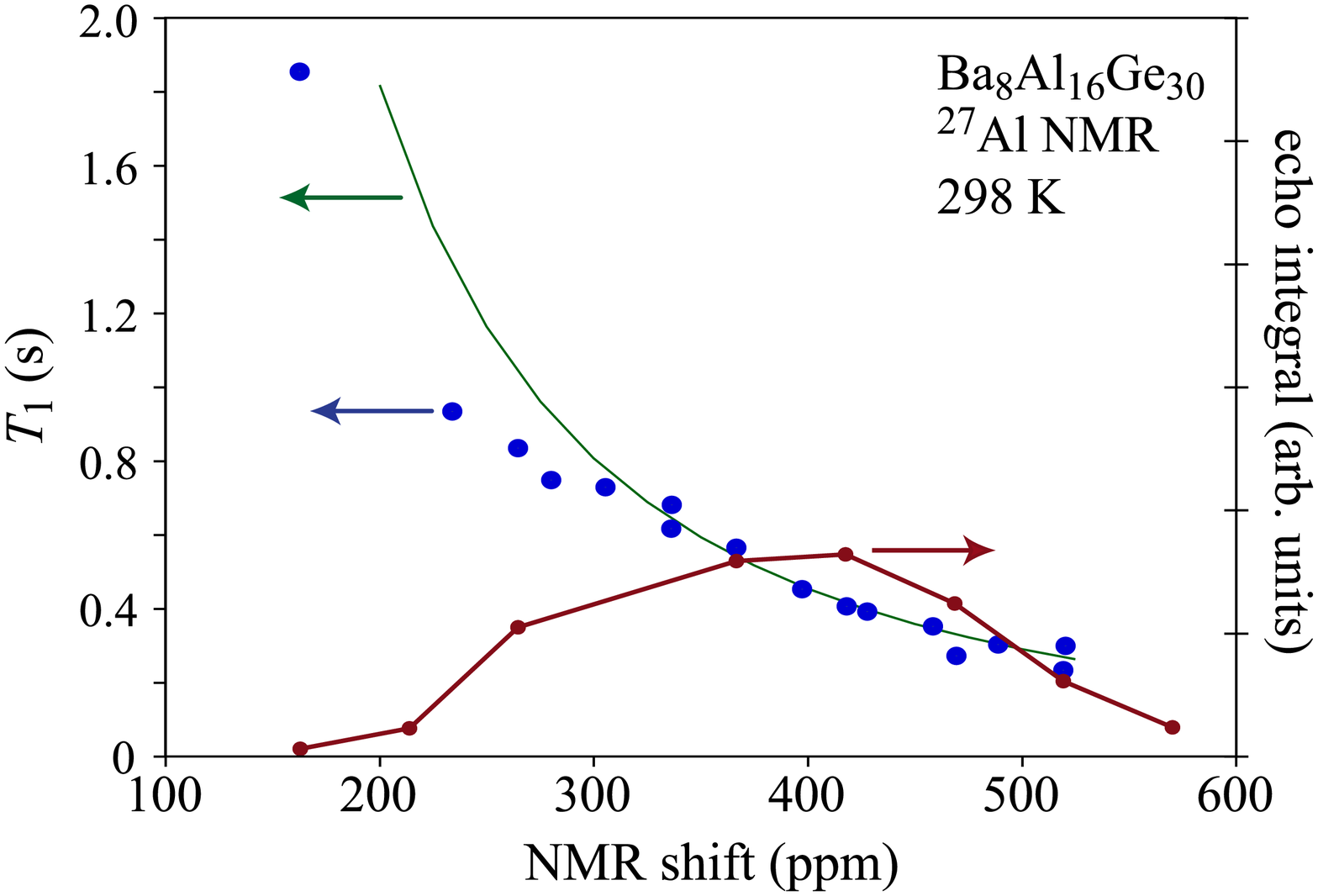}
\caption{\label{fig:fig3}(Color online) Frequency dependence of $T_{1}$ for Ba$_{8}$Al$_{16}$Ge$_{30}$ compared to central line shape (lower curve). Solid curve with monatonic decay is $T_{1}$ calculated according to Korringa law. }
\end{figure}

To further understand the central line shifts, we measured the frequency dependence of $T_{1}$, the NMR longitudinal relaxation time, at room temperature. For the 
Al$_{16}$ sample (Fig.~\ref{fig:fig2}), $T_{1}$ decreases gradually with increasing frequency, whereas for 
Al$_{12}$ (Fig.~\ref{fig:fig3}) there is a plateau coinciding with the lower-frequency peak in the central line. At the upper end both systems follow a Korringa law for $T_{1}$ vs. shift, indicating the shift to be dominated by a Knight shift ($K$). The Korringa law may be written \cite{Slichter,carter77}, 
\begin{equation}
K^{2}T_{1}T=K^{-1}(\alpha) S\equiv K^{-1}(\alpha) \frac{\hbar}{4\pi
k_B}\frac{\gamma_{e}^{2}}{\gamma_{n}^{2}},
\label{eq:KP}
\end{equation}
where $\gamma_{e}$ and $\gamma_{n}$ are the electron and nuclear gyromagnetic ratios, respectively, and $K^{-1}(\alpha)$ represents a multiplier which among other things may be due to electron-electron interactions. In simple metals it is commonly observed that $K^{-1}(\alpha)$ has a value in the range 1--5. The curves in Figs.~\ref{fig:fig2} and \ref{fig:fig3} were drawn based on Eqn.~\ref{eq:KP}, with $K^{-1}(\alpha)$ = 2.7 and 5.5, respectively. This behavior points to an inhomogeneous distribution of Knight shifts as the main factor determining the higher-frequency portions of both lines. For Al$_{12}$ the lower peak exhibits a plateau in $T_{1}$ indicating some other source of broadening for that portion of the line. Measurements on Al$_{13}$ yielded similar results to those of Al$_{12}$. The presence of an inhomogeneous distribution of isotropic shifts explains why the MAS spectra are little narrowed compared to the static spectra, since such a distribution is not narrowed through spinning.

Central line shifts also include a second-order quadrupole contribution; as shown by the simulations below, this term generally represents a few percent of the total shift for these central lines, except for the upper peak in 
Al$_{12}$ where it is somewhat larger. A chemical shift (due to $T$-independent Van Vleck susceptibility terms \cite{carter77}) may also contribute to the total shift. For comparison, the
group III-V zinc-blende semiconductors AlP, AlAs, and AlSb exhibit paramagnetic chemical shifts from 70 to 
140 ppm \cite{Sears80,Han88}. Shifts of this order of magnitude could be responsible for a portion of the Korringa enhancement identified above. 

Note that for $^{27}$Al NMR in Ba$_{8}$Al$_{14}$Si$_{31}$ \cite{Condron06}, three separated peaks could be assigned to the three framework atom types, with 6$c$ having a rather distinctive shift. That is not the case here, instead the individual sites overlap giving a line that is a superposition of framework sites. A calculation described below shows for a model configuration how this results from a large distribution of Knight shifts producing the central line shape. 

\begin{figure}
\begin{center}
\includegraphics[width=\columnwidth]{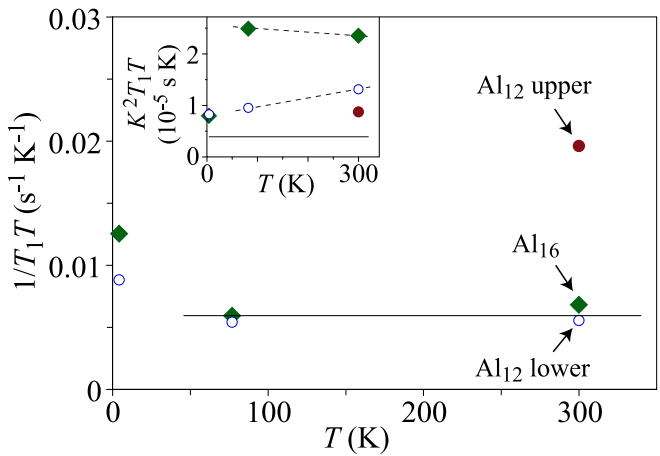}
\caption{\label{fig:fig6k2t1t} (Color online) $T_{1}T$ reciprocal product for Al$_{12}$ and Al$_{16}$ samples at peaks of the central lineshapes. Horizontal line indicates Korringa behavior. Inset:  $K^{2}T_{1}T$ product measured at the same peak positions, with same symbols as main plot. Horizontal line: Korringa relation for bare $^{27}$Al nucleus. Dashed lines are guides to the eye.}
\end{center}
\end{figure}

$(T_{1}T)^{-1}$ values determined at three temperatures, for peak positions of the Al$_{12}$ and Al$_{16}$ central lines, are shown in 
Fig.~\ref{fig:fig6k2t1t}. $(T_{1}T)^{-1}$ is constant at higher temperatures for the main parts of both lines, indicating Korringa behavior (horizontal line in the figure), but deviates upward at 4 K. The upper Al$_{12}$ peak partially merges with the lower peak and thus could not be assigned a shift at low temperatures. In the inset of 
Fig.~\ref{fig:fig6k2t1t} is shown the $K^{2}T_{1}T$ product, using for $K$ the total measured shift. With $K^{-1}(\alpha)$ = 1 indicated by the solid line, $K^{2}T_{1}T$ is enhanced for both materials as described above, and also varies slowly with temperature (dashed lines), mostly due to slow variations in $K$. Such variations in $K$ and $T_{1}$ are not unlike those reported for the host resonance in Si:P heavily doped above the metal-insulator transition, and may indicate that local composition fluctuations lead to an impurity band partially merged with the conduction band \cite{Meintjes05,Hoch05}. Similar behavior was found for $^{71}$Ga NMR in 
Ba$_{8}$Ga$_{16}$Ge$_{30}$, while in Sr$_{8}$Ga$_{16}$Ge$_{30}$ simple Korringa behavior was observed over a wide temperature range, as in a simple metal \cite{Gou05}.

\section{COMPUTATIONAL METHODS AND DISCUSSION}

{\it Ab-initio} calculations were carried out using the full-potential linearized augmented plane-wave method as implemented in the WIEN2k package \cite{blaha}, both for analysis of the electronic structure and for calculation of the electric field gradients (EFGs) \cite{Blaha85}, with subsequent numerical simulation of NMR lineshapes. In all cases we used the Generalized Gradient Approximation, with the Perdew-Burke-Ernzerhof formalism for the exchange-correlation term. We considered 33 distinct superstructures having different atomic arrangements, starting with atomic positions of the reported parent 
Ba$_{8}$Ge$_{30}$Al$_{16}$ structure \cite{Eisenmann86}. Where justified by initial results we also used WIEN2k to minimize internal structural parameters before finalizing the results.

We obtained EFG's for each Al site from the calculated all-electron charge distribution, using a method used in recent years with considerable success for analysis of NMR spectra \cite{Bastow03,Cuny08}. Relations for NMR transition frequencies between the $m^{\rm{th}}$ and $(m+1)^{\rm{th}}$ levels in a single crystal are well established \cite{carter77}, with quadrupole shifts given in terms of the EFG principal values 
($\frac{\partial ^{2}V}{\partial x ^{2}}$, $\frac{\partial ^{2}V}{\partial y ^{2}}$, $\frac{\partial ^{2}V}{\partial z ^{2}}$), 
$\eta = (\frac{\partial ^{2}V}{\partial x ^{2}}-\frac{\partial ^{2}V}{\partial y ^{2}})/\frac{\partial ^{2}V}{\partial z ^{2}}$, and 
$\nu_{Q}=\frac{3e^{2}qQ}{2I(2I-1)h}$, with $eq$ = $\frac{\partial ^{2}V}{\partial z ^{2}}$ the largest principal value and $Q$ the nuclear quadrupole moment. 
From computed EFG matrix elements the quadrupole lineshape was simulated by summing a contribution from each $\textit{m}$ to $m+1$ transition, with a 
weighting factor $\sqrt{(I+m)(I-m+1)}$ for each $\textit{m}$, and using standard expressions for the angular dependence \cite{carter77}. The results were used to calculate the powder patterns numerically, by summing uniformly over all solid angles, including the first and second-order quadrupole shift for each transition. For central lines, the only significant quadrupole contribution is the second-order term, whereas the shift of the broad background is dominated by first-order quadrupole shifts. In some cases an anisotropic chemical shift was also included in the angular summation, although this did not provide apparent improvement and hence was not used in any of the final simulations. 

The 10 configurations considered for Ba$_{8}$Ge$_{30}$Al$_{16}$ are summarized in Table~\ref{tab:table1}, where the first three columns indicate the number of Al atoms per cell in each framework site of the parent $Pm\overline{3}n$ structure, and the fourth shows the number of adjacent Al-Al pairs per cell. For these superstructures we retained the cubic cell symmetry, leaving 54 independent atoms ($P$1 space group), except for a few cases allowing higher symmetry. 8 of the configurations are the same as considered in reference 
\cite{Blake99} for Ba$_{8}$Ga$_{16}$Ge$_{30}$, while in reference \cite{Nenghabi08} a 3-4-9 configuration was considered, similar to the structure determined for Ba$_{8}$Si$_{31}$Al$_{14}$ \cite{Condron06}. 

Computed NMR lineshapes could be distinguished by the shapes of the non-central portions of the lines, and we found that Al atoms situated in Al-Al pairs contributed the largest EFG's. For configurations with 3 or more Al-Al pairs this led to broad simulated lines providing poor agreement with the data. Thus we rule out configurations with larger numbers of Al-Al pairs, a feature that is very difficult to determine via diffraction experiments.  These configurations also had the highest energies; column 5 of Table~\ref{tab:table1} shows computed total energies per formula unit after parameter optimization, relative to the most stable configurations. Note that in this case only the internal parameters were optimized, leaving lattice constants identical. The last four configurations were not optimized due to the long required computing time, however the default parameters gave poor fits and relative energies greater than 2 eV, and for other configurations optimization produced small changes in computed lineshapes and in total energy (typically 0.2 eV).

\begin{table}
\caption{\label{tab:table1} Al framework occupation used in calculations of Ba$_{8}$Al$_{16}$Ge$_{30}$, per formula unit, along with the number of Al-Al pairs per cubic cell, $\chi ^{2}$ from the lineshape simulation, and the relative calculated energies per formula unit.}
\begin{ruledtabular}
\begin{tabular}{cccccc}
6$c$ & 16$i$ & 24$k$ & Al-Al & $\chi ^{2}$ & Energy (eV)\\
\hline
3 & 1 & 12 &  0 & 2.9 & 0.00\\
3 & 3 & 10 &  0 & 2.3 & 0.60\\
3 & 4 &  9 &  1 & 8.0 & 0.00\\
3 & 4 &  9 &  2 & 1.6 & 1.03\\
4 & 2 & 10 &  2 & 5.5 & 1.28\\
5 & 3 &  8 &  4 & 6.4 & 2.31\\
4 & 6 &  6 & 3 & 3.6 & -- \\
5 & 1 & 10 &  7 & 5.4 & -- \\
6 & 4 &  6 &  5 & 4.6 & -- \\
6 & 0 & 10 & 10 & 3.9 & -- \\
\end{tabular}
\end{ruledtabular}
\end{table}

\begin{figure}
\includegraphics[width=\columnwidth]{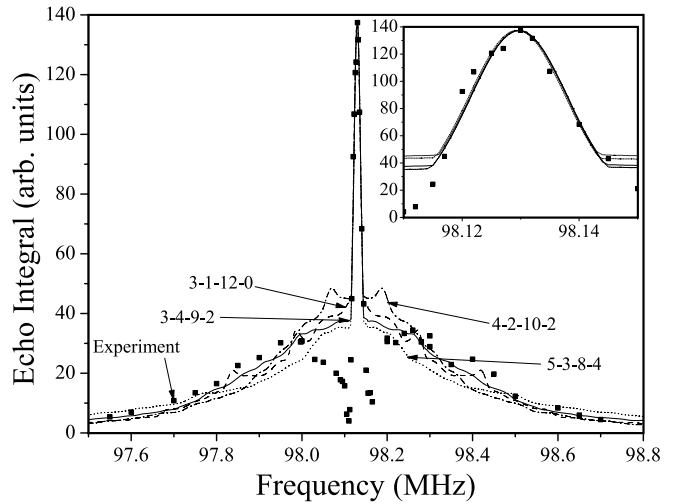}
\caption{\label{fig:fig6} Simulated NMR spectra for several Ba$_{8}$Al$_{16}$Ge$_{30}$ configurations, together with spin-echo NMR spectrum measured at 4 K. Inset: expanded view of the central-transition region.}
\end{figure}

Fig.~\ref{fig:fig6} shows several simulations compared to the Al$_{16}$ spin-echo spectrum. As in Fig.~\ref{fig:fig5}, the two valleys immediately adjacent to the central line are experimental artifacts . Among structures considered, 3-4-9-2 provided the best agreement, while the 3-3-10-0 and 3-1-12-0 also produced acceptable fits, the four numbers referring to the Al site occupations and number of Al-Al pairs, as shown left-to-right in Table~\ref{tab:table1}. In the simulations an identical distribution of isotropic shifts was added to all sites, corresponding to the breadth of the central line, and representing a Knight shift distribution as evidenced by the $T_{1}$ measurements. Additional center-of-mass shifts for each site due to the second-order quadrupole effect are relatively small; such shifts are responsible for the just-discernible frequency differences between curves in the inset of Fig.~\ref{fig:fig5}. The only other parameter used was a single overall amplitude adjustment, the same for each configuration, to make the calculated central lines match the experiment. This produced a good match for the entire lineshape, with the 3-4-9-2 simulation falling particularly close to the data. The sum of square differences between the data and each fit provide an indication of the goodness of fit, as shown by the $\chi^{2}$ (Table~\ref{tab:table1}). Note that NMR, as a local probe, is much less sensitive to widely spaced defects and surface effects that, for example, transport measurements. This can explain why the fitting based on computed results for perfect single crystals can work well for polycrystalline samples.

Although 3-4-9-2 provided the best fit, 3-1-12-0 and 3-4-9-1 had nearly equivalent total energies that were the lowest calculated for Al$_{16}$ configurations. Of these, 3-1-12-0 was also the minimum energy configuration identified in reference \cite{Blake99} for 
Ba$_{8}$Ge$_{30}$Ga$_{16}$, featuring a regular pattern of alternating Ge and group III atoms around each six-sided ring of the larger cages. Given its equivalent energy in our results, we would expect the 3-4-9-1 configuration to predominate at the growth temperature due to its greater configurational entropy. Yet, this configuration yielded a distribution of EFG's too small to agree with the NMR lineshapes. Instead, 3-4-9-2, with one additional Al-Al neighbor pair, was among the best fitting. (3-4-9-0 can also be formed, and it seems likely that its calculated energy would be lower than either 3-1-12-0 or 3-4-9-1, but presumably with a poor fit to the NMR.) The 3-4-9-$x$ configurations are in line with the general framework occupation rules established in reference \cite{Christensen07} from neutron scattering, and also consistent with our powder XRD refinement. Thus according to our results 3-4-9-2 is the best representation of the alloy configuration; it has more Al-Al pairs than energetically favored, although the difference is small and it is possible that entropic processes or incomplete equilibration may enhance the likelihood of this configuration.

\begin{figure}
\includegraphics[width=\columnwidth]{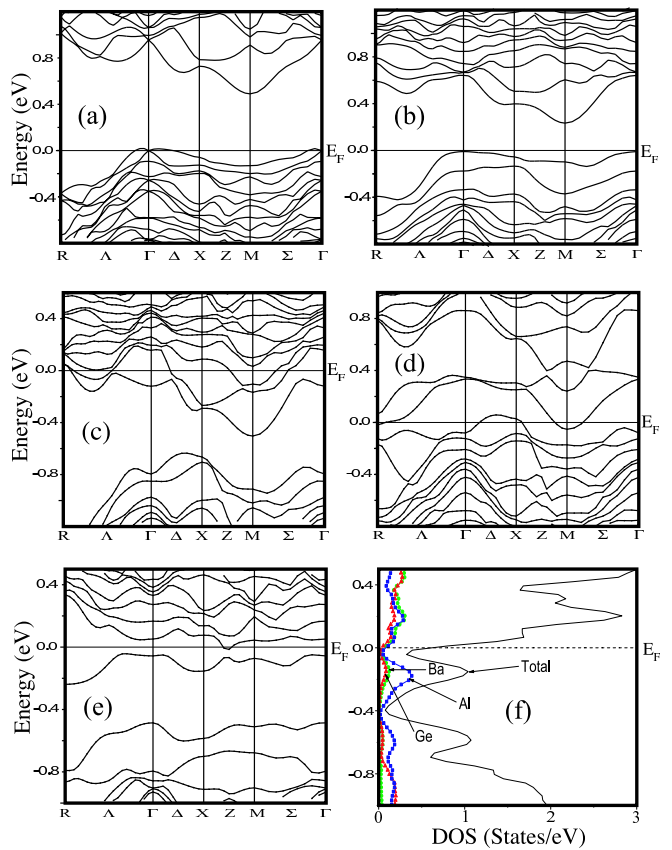}
\caption{\label{fig:bandstr_fig} (Color online) Calculated electronic structures for Ba-Al-Ge clathrate configurations. (a) and (b): 
Bandstructures for Ba$_{8}$Al$_{16}$Ge$_{30}$, 3-1-12-0 and 3-4-9-2 respectively, notation as defined in text; (c), (d), (e):
Bandstructures for Ba$_{8}$Al$_{12}$Ge$_{34-y}\Box_{y}$, $\Box_{0}$ 4-8-0-0-0, $\Box_{1}$ 4-8-0-0-0, and $\Box_1$ 0-0-12-0-2  configurations respectively. (f): Density of states for the latter, including partial densities for Al adjacent to a vacancy, Ge (6$c$), Ba (2$d$), and the total scaled by 1/10.}
\end{figure}

Bandstructures for Al$_{16}$ configurations 3-1-12-0 and 3-4-9-2 are shown in Fig.~\ref{fig:bandstr_fig} (a) and (b). Both are semiconducting, the latter in general agreement with a previously report \cite{Nenghabi08}. Our results for 3-4-9-1 and 3-4-9-2 are almost identical. The three-fold minimum at $M$ is nearly isotropic, with an effective mass 1.93 $m_{e}$  for 3-1-12-0 (2.35 $m_{e}$ for 3-4-9-2). Decomposing the calculated density at this minimum inside the muffin-tin spheres, we find average 
Al $s$-partial contributions of 0.16\%/0.18\% for 3-1-12-0/3-4-9-2 (percentages relative to the entire 54-atom cell). In the effective mass approximation, the Pauli susceptibility ($\chi_{p}$) can be calculated for a given electron density, since it is proportional to the density of states\cite{Ashcroft76}. Assuming conduction electrons introduced extrinsically through the presence of electronically active defects, we can thereby obtain the corresponding Knight shift using
\begin{equation}
K=\frac{\chi_{p}V_{cell}f_{{\rm Al}s}H_{HF}}{\mu_{B}},
\label{eq:K}
\end{equation}
where $f_{{\rm Al}s}$ is the calculated partial $s$ fraction on a given Al site, and $H_{HF}$ is the $s$-hyperfine field, taken to be 1.9 MG \cite{carter77}. Taking the average central-line shift at room temperature (0.0375\%) to be a Knight shift, the computed results yield 
$n$ = 1.1 $\times$ 10$^{21}$ cm$^{-3}$ for 3-1-12-0, corresponding to a small degenerate carrier pocket. 4.1 $\times$ 10$^{20}$ cm$^{-3}$ is calculated for 3-4-9-2. The calculated carrier densities are in the same range as those found for as-grown 
Ba$_{8}$Ga$_{16}$Ge$_{30}$ and Sr$_{8}$Ga$_{16}$Ge$_{30}$ \cite{Kuznetsov00}, typically attributed to small departures of the framework composition from the ideal semiconducting materials. Note that Eqn.~\ref{eq:K} will be valid only if enough carriers are present to reach the degenerate regime, which can be verified for the densities found here. This result is also consistent with the observed positive temperature coefficient of resistivity. The wide-line NMR fits, which are dominated by the EFG's, will be affected very little by the small density of such carriers. While these shifts are metallic, it is not unreasonable that the Zintl concept may still apply, since the carrier densities are thus the departure from the semiconducting configuration are small.

\begin{figure}
\includegraphics[width=\columnwidth]{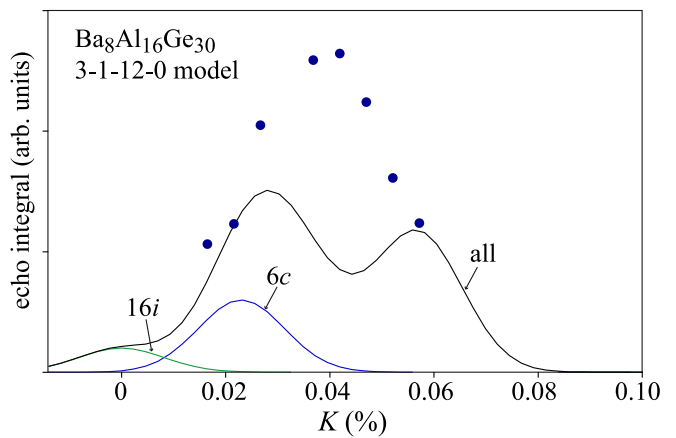}
\caption{\label{fig:Kmodel} (Color online) Calculated NMR line shape from computed Knight shifts for 3-1-12-0 configuration, with carrier density 
1.1 $\times$ 10$^{21}$ cm$^{-3}$, compared to Al$_{16}$ room-temperature data (filled circles). Calculated 6$c$ and 16$i$ portions are as indicated; remainder due to 24$k$ sites.}
\end{figure}

In addition to the average shift, the range of calculated partial densities on individual sites is large and can also account for the measured NMR line broadening. This is shown in Fig.~\ref{fig:Kmodel}, where by superposing lines for each Al site shifted according to its calculated $f_{{\rm Al}s}$ (with a single vertical scaling parameter) we obtain an inhomogeneously broadened line with a slightly larger intrinsic width than the measured central line. Results for 3-4-9-2 are similar with a larger width approximately twice that of the data. The detailed shape, with two main peaks, is not representative of the data, indicating that the calculated superstructures are only approximations to the alloy configuration, however clearly the width of the observed NMR line is an indication of the large variation of conducting states among local atomic configurations in the alloy.

In modeling the Al$_{12}$ alloy we considered the 16 superstructures summarized in 
Table~\ref{tab:table2}. The expected Al$_{12}$ Zintl composition is 
Ba$_{8}$Ge$_{31}$Al$_{12}\Box_1$, however for a nominal Al$_{10}$ composition we found more vacancies than expected in WDS scans, so we considered a range of vacancy concentrations. In all cases, vacancies were placed only on 6$c$ sites, shown to be the preferred site for vacancies in the type-I structure \cite{Fukuoka03,Budnyk04,Shimizua07}. 

\begin{table}
\caption{\label{tab:table2} Configurations considered in calculations of Ba$_{8}$Al$_{12}$Ge$_{34-y}$. Number of framework vacancies, Al framework occupations, number of Al-Al pairs and number of Al adjacent to vacancies are given per formula unit. Calculated lattice constant and relative formation energy, per formula unit of Ba$_{8}$Al$_{12}$Ge$_{34}$, given for minimized configurations.}
\begin{ruledtabular}
\begin{tabular}{cccccccc}
vacancy & 6$c$ & 16$i$ & 24$k$ &Al-Al& Al-vac & $a$ ({\AA}) & energy (eV)\\
\hline
$\Box_0$ & 4 & 8 & 0& 0 & 0 &11.01&0.00\\
$\Box_1$ & 4 & 8 & 0& 0 &  0 &11.01&0.22\\
$\Box_1$ & 4 & 4 & 4& 0 &  2 &10.99&1.85\\
$\Box_1$ & 4 & 3 &  5 &  1& 5  &&\\
$\Box_1$ & 4 & 2 &  6 &  1& 5  &&\\
$\Box_1$ & 0 & 0 &  12 &  0& 0  &10.97&1.08\\
$\Box_1$ & 0 & 0 &  12& 0 &  1  &10.98&1.08\\
$\Box_1$ & 0 & 0 &  12& 0 &  2  &10.97&1.30\\
$\Box_1$ & 0 & 0 &  12 & 1&  4  &10.98&1.98\\
$\Box_2$ & 4 & 8 & 0 &  0& 0  &10.84&1.86\\
$\Box_2$ & 4 & 4 & 4 &  0 & 2 &&\\
$\Box_2$ & 4 & 0 &  8 &  0  & 8&&\\
$\Box_2$ & 4 & 3 & 5 &  1  &5&&\\
$\Box_2$ & 4 & 2 &  6 &  1  & 6&&\\
$\Box_2$ & 0 & 0 &  12& 2 &4  &10.97&3.56\\
$\Box_3$ & 0 & 0 & 12 & 0& 6 &10.95&5.03\\
\end{tabular}
\end{ruledtabular}
\end{table}

\begin{figure}
\includegraphics[width=\columnwidth]{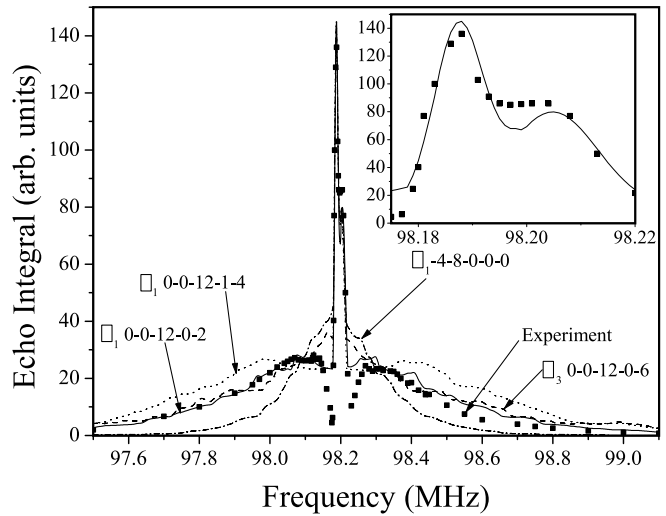}
\caption{\label{fig:fig92} Simulated NMR lines for several configurations of Ba$_{8}$Al$_{12}$Ge$_{34-y}$ together with spin-echo NMR lineshape. Inset: expanded view of the central transition region, with calculated spectrum only for 
$\Box_1$ 0-0-12-0-2 configuration with one vacancy shown.}
\end{figure}

Several calculated spectra are shown in Fig.~\ref{fig:fig92}, along with the Al$_{12}$ spin-echo spectrum. The $\Box_1$ 0-0-12-0-2 and $\Box_3$ 0-0-12-0-6 configurations (notation corresponding to first 6 columns of 
Table~\ref{tab:table2}) provided a good fit, while other configurations agreed poorly. The large difference between calculated spectra was due most significantly to the large EFG for sites adjacent to a vacancy, ranging from $\nu_{q}$ = 1000 to 2800 kHz. 
Normalization was similar to that of the 
Al$_{16}$ curves, with a single overall scaling used to force agreement in the central-line region. However, because of the two-peaked central line, two distinct isotropic shifts were applied, with a portion of the sites having a larger shift than the others, plus an additional overall broadening function applied to the otherwise relatively narrow computed central-line. The shift distribution has almost no effect on the broad non-central region of the lines, and thus does not affect the fitting except for the appearance of the central line. We will return to consider a more detailed calculation of the Knight shift distribution below.

Calculated total energies and lattice constants are shown in Table~\ref{tab:table2} for minimized configurations. In the formation energies we added the all-electron total energy per atom for diamond-structure Ge, after volume minimization using the same GGA-PBE formalism, to compensate for each vacancy. 
Configurations not minimized had simulated spectra differing significantly from the data and relative energies greater than 2.5 eV before minimization. There are a number of configurations relatively close in energy, however a general trend is seen with a preference for zero or one vacancy. Of the two configurations best fitting the NMR spectra, $\Box_3$ 0-0-12-0-6 has a considerably higher calculated energy, while $\Box_1$ 0-0-12-0-2 is energetically more favored and has the expected single-vacancy Zintl composition, thus provides a likely explanation for the data. This configuration also agrees with the trend from our XRD fits indicating Al occupation transferred to 24$k$ sites as the Al concentration is reduced.

Compared to the best-fitting $\Box_1$ 0-0-12-0-2, other configurations have lower energy:
$\Box_1$ 4-8-0-0-0 is 1.1 eV/cell lower in energy and is the most stable calculated one-vacancy configuration. Filling the vacancy with Ge yields $\Box_0$ 4-8-0-0-0 with a still lower energy, although the additional vacancy-formation entropy of the $\Box_1$ configuration may negate the small difference. Yet none of the 4-8-0-0-0 simulations agree with the NMR data; 
the $\Box_1$ simulation is shown in Fig.~\ref{fig:fig92}, and that of $\Box_0$ is similar, both having EFG's too small. 

\begin{figure}
\includegraphics[width=\columnwidth]{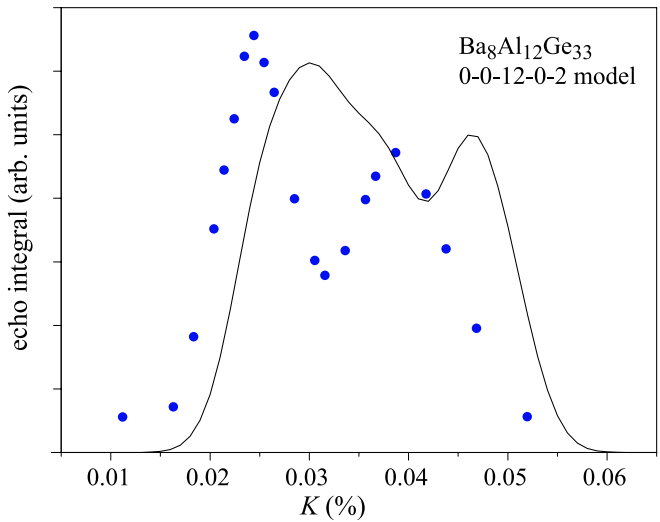}
\caption{\label{fig:KmodelAl12} (Color online) Calculated NMR line shape (solid curve) from computed Knight shifts for $\Box_1$ 0-0-12-0-2 configuration, compared to Al$_{12}$ room-temperature data (filled circles).}
\end{figure}

Band structures for some of these configurations are shown in Fig.~\ref{fig:bandstr_fig}. From the calculated densities of states for the 
$\Box_1$ 0-0-12-0-2 configuration (Fig.~\ref{fig:bandstr_fig}(f)) we can calculate the site-dependent Knight shift, using Eqn.~\ref{eq:K}, with the local susceptibility taken from the local Fermi-level density of states, using a similar method as used for the Al$_{16}$ calculation. 
The result is shown in Fig.~\ref{fig:KmodelAl12} as the solid curve, obtained by superposing a line for each of the 12 calculated Knight shifts, with an identical Gaussian broadening given to each line. Note that in contrast to the Al$_{16}$ case, for which a small carrier density was added to the otherwise insulating configuration, for this case there are no adjustable parameters, save for a vertical scaling. The calculated peak at the right edge is due to the two Al sites adjoining a vacancy in this configuration, for which the local charge density is larger. Overall, the calculated shifts are larger than observed, although it is possible that the pseudogap at $E_{F}$ might become deeper, and hence the shifts smaller, if not for the well-known tendency for band-gap underestimation in the PBE-GGA method. Thus the agreement seems quite reasonable.

While the calculated $\Box_1$ 0-0-12-0-2 configuration is not semiconducting, $E_{F}$ falls between bands over much of $k$ space (Fig.~\ref{fig:bandstr_fig}). In this way, this configuration might be considered to satisfy the Zintl criterion, in which a semiconductor would be expected, with a hybridization gap at $E_{F}$. From the partial density of states [Fig.~\ref{fig:bandstr_fig}(f)] it can be seen that contributions from the two Al vacancy neighbors dominate the highest filled band, making up the peak just below 
$E_{F}$, far outweighing contributions from any other site; the sites with the largest Ge and Ba contributions in this region give the two additional curves also plotted in Fig.~\ref{fig:bandstr_fig}(f). It is thus apparent that the associated states are largely non-bonding hybrids adjoining the vacancy. The concentration of such states for this band explains the larger Knight shift for such Al sites, and hence the observed split-off central peak.

Having minimized the internal parameters for a number of configurations, we can also address the atomic displacements which are of potential significance for thermoelectric applications. In Al$_{16}$, calculated Ba displacements from symmetry positions are very small even for 6$d$ atoms identified through diffraction experiments as having distorted nuclear density profiles \cite{Christensen07}, and which exhibit a low-energy "rattling" mode \cite{Nenghabi08}. In the 3-4-9-2 configuration, the largest calculated Ba 6$d$ displacement is 0.019 {\AA}. Of the six optimized Al$_{16}$ structures, no Ba 6$d$ was 
more than 0.07 {\AA}
from the cage center, and the overall average displacement was 0.025 {\AA}. Also the longest framework bonds appeared in structures having Al-Al neighbors, for example in the 3-4-9-2 configuration the Al-Al bonds increased to 2.61--2.77 {\AA}, compared to the Ge-Ge bond lengths of 
2.51--2.53 {\AA}; this reflects the weakness of the Al-Al bond. Calculated Ba 6$d$ displacements are consistently smaller that the experimental 
values \cite{Christensen07}. The calculated results do not include dynamical terms, and also may reflect a balance between opposing tendencies for different growth conditions, for which displacements tended toward perpendicular crystal directions \cite{Christensen07}, presumably due to different dominant defect types. For Al$_{12}$ configurations with vacancies, the calculated structural distortions are considerably larger; all framework atoms next to the vacancy shift toward the open space (by 0.35 to 0.55 {\AA}), while the four Ba 6$d$ sites in cages including the vacancy shift toward the 24$k$ framework sites (Ge or Al) immediately adjacent to the vacancy. This decreases the Ba 6$d$-(Ge/Al)24$k$ bond length from 3.64 {\AA} to 3.39--3.43 {\AA}. This is not quite as short as the sum of the covalent radii (3.35--3.36 {\AA}), however it is clear what develops is a much more localized covalent bond involving the atoms adjacent to the vacancy.

\section{CONCLUSIONS}

Wide-line $^{27}$Al NMR measurements on Ba$_{8}$Al$_{x}$Ge$_{46-y}$ clathrates yielded quadrupole-broadened lineshapes that help to establish the local configuration of Al atoms in these alloys. Combined with {\it ab initio} calculations of a number of superstructure configurations, we obtained a good fit to the NMR measurements for $x$ = 16 that corresponded to the most stable configurations identified by our computations. Analysis of the NMR and computational results showed that a large distribution of metallic shifts dominates the central region of the NMR spectrum, indicating a considerable variation of carrier density within the cell. For $x <$ 16 we found a splitting in the main NMR peak, coinciding with the appearance of vacancies in order to approximate a Zintl composition. Detailed analysis for $x$ = 12 yielded a configuration giving very good agreement with the NMR spectrum. The fitting indicates that Al atoms tend to shift to the 24$k$ framework site for $x <$ 16. From an analysis of the NMR shifts we deduced that Al non-bonding states play a significant role near the Fermi edge. Calculations also indicate rather small static distortions from the type-I structure for intrinsic Ba$_{8}$Al$_{16}$Ge$_{30}$, but more significant distortions with vacancies present.

\section{acknowledgments}

\begin{acknowledgments}
This work was supported by the Robert A. Welch Foundation, Grant
No. A-1526, by the National Science Foundation (DMR-0103455).
\end{acknowledgments}

\bibliography{Al_clathrate}

\end{document}